\documentclass[aps,prl,nofootinbib,twocolumn,superscriptaddress,preprintnumbers,nofootinbib]{revtex4-1}

\usepackage{amssymb}
\usepackage{amsmath}
\usepackage{epsfig}
\usepackage{hyperref}
\usepackage{breakurl}
\usepackage{color}

\usepackage{graphicx}
\usepackage[export]{adjustbox}

\makeatletter
\def\simgt{\mathrel{\lower2.5pt\vbox{\lineskip=0pt\baselineskip=0pt
           \hbox{$>$}\hbox{$\sim$}}}}
\def\simlt{\mathrel{\lower2.5pt\vbox{\lineskip=0pt\baselineskip=0pt
           \hbox{$<$}\hbox{$\sim$}}}}
\makeatother

\newcommand{\be}{\begin{equation}}
\newcommand{\ee}{\end{equation}}
\newcommand{\bea}{\begin{eqnarray}}
\newcommand{\eea}{\end{eqnarray}}
\newcommand{\eq}[2]{\be\begin{aligned}#1 \label{#2}\end{aligned}\ee}

\newcommand{\Eq}[1]{Eq.~\eqref{#1}}

\def\Section#1{\vskip .05 cm \noindent {\it #1}}

\begin{document}

\title{Scattering Amplitudes and the Navier-Stokes Equation}

\author{Clifford Cheung}
\affiliation{Walter Burke Institute for Theoretical Physics,
California Institute of Technology, Pasadena, CA 91125}

\author{James Mangan}
\affiliation{Walter Burke Institute for Theoretical Physics,
California Institute of Technology, Pasadena, CA 91125}

\begin{abstract}

We explore the scattering amplitudes of fluid quanta described by the Navier-Stokes equation and its non-Abelian generalization.  
These amplitudes exhibit universal infrared structures analogous to the Weinberg soft theorem and the Adler zero.  
Furthermore, they satisfy on-shell recursion relations which together with the three-point scattering amplitude furnish a pure S-matrix formulation of incompressible fluid mechanics.  Remarkably, the amplitudes of the non-Abelian Navier-Stokes equation also exhibit color-kinematics duality as an off-shell symmetry, for which the associated kinematic algebra is literally the algebra of spatial diffeomorphisms.  Applying the double copy prescription, we then arrive at a new theory of a tensor bi-fluid.  Finally, we present monopole solutions of the non-Abelian and tensor Navier-Stokes equations and observe a classical double copy structure.



\end{abstract}

\preprint{CALT-TH 2020-044}

\maketitle

\Section{Introduction.}  The Navier-Stokes equation (NSE) is remarkably simple and follows trivially from the laws of classical mechanics. Still, its unassuming form and humble origins belie a daunting complexity: the problem of turbulence, which has confounded physicists for generations.  The root of this difficulty is that the turbulent regime is essentially a strong coupling limit of the theory.

Of course, non-perturbative dynamics are not intractable per se.  But in prominent examples such as quantum chromodynamics, progress has hinged crucially on the existence of an action formulation.
However, there is no action whose classical extremization yields the NSE, simply because the least action principle is time reversal invariant while the viscous dynamics of a fluid are not.

Notably, the very premise of the modern S-matrix program (see \cite{ElvangHuang, DixonReview, TASIReview} for reviews) is to bootstrap scattering dynamics from first principles {\it without} the aid of an action.  These efforts have centered primarily on gauge theory and gravity, which are stringently constrained by fundamental properties like Poincare invariance, unitarity, and locality.  These theories are ``on-shell constructible'' since their S-matrices are fully dictated at tree level by on-shell recursion \cite{BCF, BCFW} and at loop level by generalized unitarity \cite{UnitarityReview}.  Remarkably, the modern S-matrix approach has also uncovered genuinely new structures within quantum field theory such as color-kinematics duality \cite{BCJ1, BCJ2}, the scattering equations \cite{CHY1, CHY2, CHY3, CHY4}, and reformulations of amplitudes as volumes of abstract polytopes \cite{Amplituhedron1, Amplituhedron2, Associahedron}. 

The NSE does not originate from an action but it nevertheless encodes an S-matrix characterizing the scattering of fluid quanta. In particular, by solving the NSE in the presence of an arbitrary source one obtains the generating functional for all tree-level scattering amplitudes \cite{Wyld}.   The turbulent regime then corresponds to the S-matrix at strong coupling, which here is unrelated to a breakdown of the $\hbar$ expansion because the NSE is intrinsically classical and hence devoid of any a priori notion of loops.\footnote{A notion of loops emerges if we introduce stochastic correlations between sources but we will not consider this here.}  Instead, turbulence is encoded in tree-level scattering processes at arbitrarily high multiplicity, where traditional perturbative methods are rather limited.   Nevertheless, there are reasons for optimism in light of the modern S-matrix program, whose tools have uncovered analytic formulae for precisely this kind of arbitrary-multiplicity process involving maximally helicity violating gluons \cite{ParkeTaylor} and gravitons \cite{Hodges}.






In this paper we initiate a study of the perturbative scattering amplitudes of the NSE and its natural non-Abelian generalization, which we dub the non-Abelian Navier-Stokes equation (NNSE).  
To begin, we recapitulate the explicit connection between equations of motion and S-matrices \cite{Wyld}, drawing on the close analogy between the incompressibility of a fluid and the transverse conditions of a gauge theory.  We present the Feynman rules for these theories and compute their three- and four-point scattering amplitudes.
Next, we examine the infrared properties of these theories, proving that they exhibit a leading soft theorem essentially identical to that of gauge theory \cite{WeinbergSoft} as well as a soft Adler zero \cite{Adler} reminiscent of the non-linear sigma model.  
Exploiting these properties, we then derive on-shell recursion relations that express all higher-point amplitudes as sums of products of three-point amplitudes, thus establishing that the NSE and the NNSE are on-shell constructible.

Remarkably, we discover that the {\it off-shell} Feynman diagrams of the NNSE automatically satisfy the kinematic Jacobi identities required for color-kinematics duality \cite{BCJ1, BCJ2}.  This implies the existence of an off-shell color-kinematic symmetry and a corresponding conservation law, which we derive explicitly.  Applying the double copy prescription, we then square the NNSE to obtain a tensor Navier-Stokes equation (TNSE) describing the dynamics of a bi-fluid degree of freedom.   Last but not least, we derive monopole solutions to the NNSE and the TNSE and discuss the classical double copy.



\medskip
\Section{Setup.}   To begin, let us consider an incompressible fluid described by a velocity field $u_i$.\footnote{Late lower-case Latin indices $i,j,k,\ldots$ run over spatial dimensions, early lower-case Latin indices $a,b,c,\ldots$ run over colors, and upper-case Latin indices $A,B,C,\ldots$ run over external legs.  Dot products are denoted by $v_i w_i = vw$ and $v_i v_i=v^2$.}  Incompressibility implies that velocity field is solenoidal, so $\partial_i u_i =0$.  The dynamics of the fluid are governed by the NSE,
\eq{
\left(\partial_0 - \nu \partial^2 \right) u_i + u_j \partial_j u_i +  \partial_i \left(\frac{p}{\rho}\right) = J_i \, ,
}{eq:NS1}
where $\rho$ is the constant energy density, $p$ is the pressure, $\nu$ is the viscosity, and $J_i$ is a source term which we also assume to be solenoidal.  Taking the divergence of \Eq{eq:NS1}, we obtain
$\partial^2 (p/\rho) = - \partial_i u_j \partial_j u_i $, from which we then solve for $p/\rho$ and insert back into \Eq{eq:NS1} to obtain
\eq{
\left(\partial_0 - \nu \partial^2 \right) u_i +\left(\delta_{ij} - \frac{\partial_i \partial_j}{\partial^2} \right)  u_k \partial_k u_j = J_i \, .
}{eq:NS2}
Hence, the pressure has the sole purpose of projecting out all but the solenoidal modes.

We can generalize this setup to an incompressible {\it non-Abelian} fluid described by a velocity field $u_i^a$ satisfying the solenoidal condition $\partial_i u_i^a=0$ and the NNSE,\footnote{The quark-gluon plasma is also described by a colored fluid \cite{Bistrovic:2002jx}, though crucially with equations of motion different from ours.}
\eq{
\left(\partial_0 - \nu \partial^2 \right) u_i^a +f^{abc}  u_j^b \partial_j u_i^c + \partial_i \left(\frac{p^a}{\rho}\right)= J_i^a \, .
}{eq:NANS1}
Here $f^{abc}$ is a fully antisymmetric structure constant and we have introduced non-Abelian versions of the pressure $p^a$ and the solenoidal source term $J_i^a$.  The divergence of \Eq{eq:NANS1}, 
$\partial^2 (p^a/\rho) = - f^{abc} \partial_i u_j^b \partial_j u_i^c  = 0$ is identically zero due to antisymmetry of the structure constants.    We thus drop the pressure altogether to obtain 
\eq{
\left(\partial_0 - \nu \partial^2 \right) u_i^a +f^{abc}   u_j^b \partial_j u_i^c   = J_i^a \, ,
}{eq:NANS2}
The absence of a projector in \Eq{eq:NANS2} as compared to \Eq{eq:NS2} results in substantial simplifications.

The NSE is simply conservation of energy-momentum, $\partial_0 T_{0j} = \partial_i T_{ij} $, in the Newtonian limit where $T_{0i} =-\rho u_i $ and $T_{ij} =\rho u_i u_j + p \delta_{ij} - \rho \nu \partial_{(i} u_{j)}$.  Analogously, the NNSE can be recast as conservation of a peculiar non-Abelian tensor, $\partial_0 T_{0j}^a = \partial_i T_{ij}^a $, where $T_{0i}^a =-\rho u_i^a $ and $T_{ij}^a =\rho f^{abc} u_i^b u_j^c + p^a \delta_{ij} - \rho \nu \partial_{(i} u_{j)}^a $.

\medskip
\Section{Amplitudes.}  As is well-known, the tree-level S-matrix can be computed by solving the classical equations of motion for a field in the presence of arbitrary sources.  The field itself is the generating functional of all tree-level scattering amplitudes.  Hence, the Berends-Giele recursion relations for gauge theory \cite{BerendsGiele} and gravity \cite{CliffSimplicityGR} are literally the classical equations of motion.  Applying identical logic to the NSE, one obtains the Wyld formulation of fluid dynamics \cite{Wyld}, which we summarize below.

To begin, we define the notion of an ``asymptotic'' quantum of fluid.  Inserting a plane wave ansatz $u_i \sim \varepsilon_i e^{-i\omega t} e^{ipx}$ into the linearized NSE and the solenoidal condition, we obtain the on-shell conditions,  $i\omega - \nu p^2 =0$ and $ p \varepsilon=0$,
which exactly mirror those of gauge theory.  The solenoidal condition eliminates the longitudinal mode, leaving $-$ and $+$ helicity modes corresponding to left and right circularly polarized fluid quanta.
Since the on-shell energy is imaginary, the on-shell solution, $u_i \sim \varepsilon_i e^{-\nu p^2 t} e^{ipx}$, is a diffusing wavepacket, as expected for a fluid velocity field undergoing viscous dissipation.


Next, we solve the NSE perturbatively in the source to obtain the one-point function of the velocity field $u_i(t,x,J)$ as a function of spacetime and a functional of $J$.  Fourier transforming to energy and momentum space yields $u_i(\omega, p, J)$, whose functional derivative $G_{n+1}=\left[\prod_{A=1}^n \varepsilon_{Ai_A}  \,\frac{\delta }{\delta J_{i_A}(\omega_A, p_A)}\right] u_i(\omega, p,J) \big|_{J=0}$
is the correlation function for $n$ fluid quanta which are emitted by $J$ and subsequently absorbed by the one-point function, $u_i(\omega, p, J)$.  Here $(\omega_A,p_A)$ are the energy and momentum flowing from each ``leaf'' leg originating from an emission and $(\omega,p) =( \sum_{A=1}^n \omega_A,  \sum_{A=1}^n p_A)$ are the total energy and  momentum flowing into the ``root'' leg upon absorption.      The scattering amplitude $A_{n+1}$ is then obtained from $G_{n+1}$ by amputating the external legs and stripping off the delta functions for energy and momentum conservation.  For the remainder of this paper we assume that the leaf legs are on-shell but the root leg is not, so $A_{n+1}$ is in actuality a {\it semi-on-shell} amplitude.

The Feynman rules for the NSE can be found in \cite{Wyld} so we do not present them again here.   Instead we focus on the NNSE.  The propagator in this theory is
\eq{
\raisebox{0ex}{\includegraphics[trim={0 0 0 0},clip,valign=c]{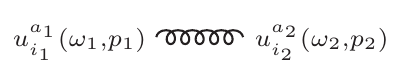}} =  \frac{  \delta^{a_1 a_2}\delta_{i_1 i_2}}{i\omega_1 - \nu p_1^2} \, .
}{eq:prop}
where the energy flow direction is important for the sign of $\omega_1$.
The only interaction is the three-point vertex,
\eq{
\hspace{-1pt}\raisebox{0.5ex}{\includegraphics[trim={0 0 0 0},clip,valign=c]{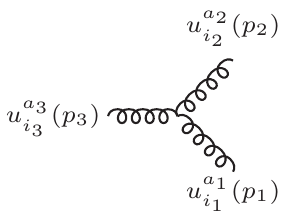}}  \begin{array}{c}
=   f^{a_1 a_2 a_3} \left( p_{1 i_2} \delta_{ i_1 i_3} - p_{2 i_1} \delta_{i_2 i_3}    \right)   \\ \\
\sim  f^{a_1 a_2 a_3} (p_{3 i_1} \delta_{i_2 i_3} -p_{3 i_2} \delta_{i_1 i_3} )
\end{array} \, ,
}{eq:V3}
where in the second line we have used momentum conservation together with the fact that all terms proportional to $p_{1 i_1}$ or $p_{2 i_2}$ vanish when dotted into sources or interaction vertices due to the solenoidal condition.  Note that the kinematic factors in \Eq{eq:V3} are not fully antisymmetric since the root leg and the leaf legs are distinguishable.
The Feynman rules for NSE are identical except with color structures dropped and plus signs in \Eq{eq:V3}.

Remarkably, the above Feynman rules imply that all amplitudes are manifestly {\it energy independent} in the sense they they depend only on dot products of momenta and polarizations.  This is property is obvious for the three-point interaction vertex but slightly more subtle for the propagators.  Regarding the latter, consider that the energy and momentum of each leaf leg is  $(\omega_A, p_A)$, so the energy and momentum flowing through any intermediate leg is $(\sum_A \omega_A , \sum_A p_A)$, where the sum runs over a subset of the legs.  The corresponding propagator is then
\eq{
\frac{ \delta^{a_1 a_2}\delta_{i_1 i_2}}{ i \sum\limits_A \omega_A - \nu \big(\sum\limits_A p_A \big)^2} =  - \frac{1}{\nu} \frac{ \delta^{a_1 a_2}\delta_{i_1 i_2}}{  \sum\limits_{A\neq B} p_A p_B} \, ,
}{eq:props}
which is independent of energy.  Here we have made use of the on-shell conditions for the leaf legs.   

From \Eq{eq:props} we realize that each propagator appears with the effective coupling constant $1/\nu$, in perfect analogy with graviton perturbation theory, where each propagator appears with the gravitational constant $G$.  Hence, the turbulent regime of high Reynolds number, i.e.~low viscosity, corresponds to strong coupling.

Let us consider a few examples.  The three-point scattering amplitude of the NNSE is
\eq{
A(123) &= f^{a_1 a_2 a_3} \times \Big[ (p_1 \varepsilon_2) (\varepsilon_1 \varepsilon_3) - \{ 1\leftrightarrow 2 \} \Big] \, ,
}{eq:A3}
while the four-point scattering amplitude is
\begin{widetext}
\eq{
A(1234) &= f^{a_1 a_2 b}f^{b a_3 a_4} \times \frac{1}{p_1 p_2} \Big[(p_1 \varepsilon_2)(p_3 \varepsilon_1) (\varepsilon_3 \varepsilon_4) + (p_1 \varepsilon_2)(p_4 \varepsilon_3) (\varepsilon_1\varepsilon_4)  - \{ 1\leftrightarrow 2 \} \Big]+ t\textrm{-channel}+u\textrm{-channel} \, ,
}{eq:A4}
\end{widetext}
dropping all coupling constant prefactors $\nu$ throughout.  As advertised there is no explicit energy dependence.

It is natural to translate these amplitudes into the non-relativistic spinor helicity formalism of \cite{Maldacena}.  However, the usual simplifications enjoyed by relativistic gauge theories do not occur here, for two reasons.  First, non-relativistic kinematics permit new inner products of angle and square spinors which are rotationally invariant but not Lorentz invariant.  For example, the NNSE three-point scattering amplitude in \Eq{eq:A3} becomes
\eq{
A(1^- 2^- 3^-) &= k \, \langle 12 \rangle \langle 23 \rangle  \langle 3 1\rangle\\
A(1^+ 2^- 3^-) &= k \, [1 2\rangle \langle 23\rangle \langle 31] \\
A(1^- 2^- 3^+)&= k \, [31\rangle \langle 12 \rangle \langle 23]
}{eq:A3_spinors}
where $k= f^{a_1 a_2 a_3}/\langle 1 1 ] \langle 2 2]$ and with all other helicity configurations obtained by conjugation or permutation.  Notably, \Eq{eq:A3_spinors} can be recast into the form of gauge theory amplitudes multiplying a Lorentz violating, purely energy-dependent form factor, e.g.~$A(1^-2^-3^+) =  f^{a_1 a_2 a_3} \frac{\langle 12\rangle^3}{\langle 13 \rangle \langle 32\rangle }  \times  
\left[ 1-\frac{\langle 1 1]}{\langle 2 2]}-\frac{\langle 2  2]}{\langle 1 1]} \right]$.
So little group covariance constrains the amplitudes to an extent but there is substantial freedom left due to the breaking of Lorentz invariance.
The second reason spinor helicity formalism does not simplify expressions is that the theory does not exhibit helicity selection rules, as is clear from \Eq{eq:A3_spinors}.  Hence the helicity violating sectors of the theory are not simpler than others, in contrast with gauge theory.

\medskip
\Section{Soft Theorems.} We now turn to the infrared properties of these amplitudes.  
First, note that the NSE and NNSE amplitudes trivially exhibit an Adler zero,  
\eq{
\underset{p\rightarrow 0}{\lim} \, A_{n+1}(p_1,\cdots , p_n) = 0 \, ,
}{eq:Adler}
when momentum of the root leg, $p = \sum_{A=1}^n p_A$, is taken soft.
This is property is obvious from the Feynman rule for the three-point interaction vertex, e.g.~as shown in \Eq{eq:V3} for the NNSE, which is manifestly proportional to the momentum of the root leg.  Physically, the Adler zero arises because the NSE and NNSE are in fact conservation equations, $\partial_0 T_{0j} = \partial_i T_{ij} $ and $\partial_0 T_{0j}^a = \partial_i T_{ij}^a $, for which every term has a manifest derivative.


Second, every leaf leg of a NSE amplitude satisfies a universal leading soft theorem,
\eq{
\underset{p_n \rightarrow 0}{\lim} \, A_{n+1}(p_1,\cdots, p_n)=  \left[ \, \sum_{A=1}^{n-1} \frac{ p_A e_n}{ p_A p_n} \,  \right]A_n(p_1, \cdots, p_{n-1}) \, ,
}{eq:softpole}
which is highly reminiscent of the Weinberg soft theorem in gauge theory \cite{WeinbergSoft}.
To derive \Eq{eq:softpole} we realize that the most singular contribution in the $p_n\rightarrow 0$ limit arises when leaf leg $n$ fuses with another leaf leg $A$, resulting in a pole from the merged propagator.  The corresponding three-point vertex and propagator is
\eq{
 \underset{p_n \rightarrow 0}{\lim} \,
 \frac{1}{p_A p_n}\Big[ (p_A e_n )e_{Ai} + (p_n e_A )e_{ni}  \Big] &= \frac{p_A e_n }{p_A p_n } e_{Ai},
}{}
where the free index on the  polarization dots into a lower-point amplitude, thus establishing \Eq{eq:softpole}.  This same logic applies trivially to the NNSE as well.

Note that the NSE and NNSE do not have collinear singularities.  The two-particle factorization poles go as $1/p_Ap_B$, so there  are instead ``perpendicular'' singularities when the momenta are orthogonal.

\medskip
\Section{Recursion Relations.}   On-shell recursion is implemented by applying a $z$-dependent deformation of the external kinematics in order to generate a family of amplitudes $A_{n+1} (z)$.
The target amplitude is then $A_{n+1}(0) = \oint \tfrac{dz}{z} A_{n+1}(z)$ for a contour encircling the origin.  By Cauchy's theorem the integral can be rewritten as a sum over residues at the poles of $A_{n+1}(z)$, together with a boundary term at $z =\infty$.  If the latter is zero then the former defines an on-shell recursion relation that recasts the original amplitude in terms of sums of products of on-shell lower-point amplitudes arising from factorization poles.  In gauge theory and gravity, the boundary terms vanish for an appropriate momentum shift \cite{BCF, BCFW, NimaJared}.  More generally, the boundary term is not zero but can be avoided, provided additional knowledge about the amplitude.  For example, if the amplitude has Adler zeros \cite{Adler}, then $A_{n+1}(z_{i*})=0$ whenever $z =z_{i*}$ coincides with a soft limit of the external kinematics.  The target amplitude is then computed via $A_{n+1}(0) = \oint \tfrac{dz}{z} \prod_i \tfrac{1}{1-z/z_{i*}} A_{n+1}(z)$, which for enough Adler zeros will eliminate the boundary term without introducing new residues in the recursion relation.

Next, let us derive on-shell recursion relations for the NSE and the NNSE.  It will suffice to identify a shift of the external kinematics which either has vanishing boundary term or probes an Adler zero of the amplitude.   We divide our discussion based on whether the shift modifies the energies in the amplitude or not.

For unshifted energies the momenta must shift so as to maintain the on-shell conditions.  A natural choice is
\eq{
p_A &\rightarrow p_A + z \tau_A \varepsilon_A  \, ,
}{eq:shift1}
for the leaf legs, keeping the energies $\omega_A$ and polarizations $\varepsilon_A$ unchanged.   Here the constants $\tau_A$ are a priori unconstrained since we  implicitly shift the momentum of the rooted leg, which is off-shell, to to conserve momentum.  Note that \Eq{eq:shift1} maintains the on-shell conditions since $p_A \varepsilon_A= \varepsilon_A^2=0$ for circular polarizations.

The boundary term is obtained from the large $z$ behavior of the NSE and NNSE amplitudes.  From the Feynman rules it is obvious that these take the schematic form
\eq{
A_{n+1} \sim \sum  \frac{(p\varepsilon)^{n-1} (\varepsilon \varepsilon_{n+1})}{(pp)^{n-2}} \, ,
}{eq:A_schematic}
so every term is proportional to a single dot product of a polarization of a leaf leg with that of the root leg.   This of course has the form of the cubic Feynman diagrams of gauge theory.  Now in the large $z$ limit of \Eq{eq:shift1}, we find that $pp \sim z^2$, $p\varepsilon\sim z$, $\varepsilon \varepsilon\sim 1$, so \Eq{eq:A_schematic} implies that $A_{n+1 }\sim z^{-n+3}$.  The boundary term vanishes when $n\geq 4$, so all amplitudes at five-point and higher are constructible via this shift.  A downside of this shift is that the intermediate propagators are quadratic polynomials in $z$ so each factorization channel enters via a pair of residues in the recursion relation \cite{EFTRecursion, GenericQFTRecursion}.

Conveniently, for appropriately chosen $\tau_A$, the momentum shift in \Eq{eq:shift1} will probe the Adler zero of the root leg.  For example, at four-point we can write the momentum of the root leg as $p = \tau_1 \varepsilon_1 + \tau_2 \varepsilon_2+ \tau_3 \varepsilon_3$, so $z=1$ corresponds to the soft limit for which $A_{3+1}(1)=0$.    Hence we can compute the four-point amplitude via $A_{3+1}(0) = \int \tfrac{dz}{z}\tfrac{1}{1-z} A_{3+1}(z)$, where the factor of $(1-z)^{-1}$ improves the large $z$ convergence of the integral.

A more elegant recursion relation can be constructed if we also shift energies.  
Consider a shift of the leaf legs reminiscent of the Risager deformation \cite{Risager},
\eq{
p_A \rightarrow p_A + z (p_A \eta) \eta , \qquad
\varepsilon_A \rightarrow \varepsilon_A - z (\varepsilon_A \eta) \eta  ,
}{eq:shift2}
where $\eta^2=0$ is nilpotent and orthogonal to the polarization of the root leg, so $\eta \varepsilon_{n+1}=0$.  This guarantees that $p_A \varepsilon_A=0$ holds after the shift.  Here we also implicitly shift the energies of the leaf legs $\omega_A$ in whatever way is needed to maintain the on-shell conditions.  
At large $z$, the invariants scale as $pp \sim z$, $p\varepsilon\sim 1$, $\varepsilon \varepsilon\sim 1$, so \Eq{eq:A_schematic} implies $A_{n+1 }\sim z^{-n+2}$.   The boundary term vanishes for $n \geq 3$, so recursion applies at four point and higher.

However, on closer inspection one realizes that the Feynman diagrammatic numerators are all invariant under \Eq{eq:shift2} and so the only $z$ dependence enters through simple poles in the intermediate propagators.   As a result, the factorization diagrams that appear in the recursion relation are literally Feynman diagrams and hence not very useful.  Nevertheless it is amusing that the Feynman diagram expansion of the NSE and the NNSE is precisely analogous to the maximally helicity violating vertex expansion of gauge theory \cite{CSW}.

\medskip
\Section{Color-Kinematics Duality.}  The NNSE is purely trivalent and has a strong resemblance to gauge theory.  It is then perhaps unsurprising that it  also exhibits color-kinematics duality.  To see why, consider a triplet of {\it off-shell} Feynman diagrams describing the $s$, $t$, and $u$ channel exchange of a quantum of fluid within some larger arbitrary scattering process.  The sum of these contributions is $\tfrac{c_s n_s}{s}+\tfrac{c_t n_t}{t}+\tfrac{c_u n_u}{u}$, where $c_s, c_t, c_u$ and $n_s, n_t, n_u$ are the color factor and kinematic numerator of each Feynman diagram and $c_s + c_t + c_u=0$.  In the $s$-channel we find that $c_s = f^{a_1 a_2 b}f^{b a_3 a_4}$ while
\eq{
n_s &=\left[ p_{1 i_2} (p_{1i_3}+p_{2 i_3}) \delta_{i_1 i_4}  +p_{2i_1}p_{3i_2}\delta_{i_3 i_4}  - \{ 1\leftrightarrow 2\}\right]  \, ,
}{eq:ns}
with the $t$- and $u$-channel contributions related by permuting legs 1,2,3.  To obtain \Eq{eq:ns} we have set $p_{1i_1}= p_{2 i_2} =p_{3i_3}=0$ due to the solenoidal condition for the one-point function of the velocity field.  Remarkably, \Eq{eq:ns} implies that $n_s + n_t + n_u = 0$,
establishing an off-shell duality between color and kinematics for the NNSE.  With this in mind we recast the NNSE in \Eq{eq:NANS2} into a more suggestive form
\eq{
\left(\partial_0 - \nu \partial^2 \right) u_i^a + \frac{1}{2} f^{abc} f_{ijk}   u_j^b u_k^c   = J_i^a \, ,
}{eq:CK_EOM}
where we have defined a kinematic structure constant $f_{ijk}$ which acts as a differential operator
\eq{
f_{ijk} v_j w_k &=  v_j \partial_j w_i - w_j  \partial_j v_i  \, ,
}{}
and by construction coincides with the Feynman rule for the three-point interaction vertex.  Observing that
\eq{
\left[ v_j \partial_j , w_k \partial_k\right] = f_{ijk} v_j w_k \partial_i \, ,
}{}
we see that $f_{ijk}$ are the structure constants of the diffeomorphism algebra.  Note the similarity of this kinematic algebra to that of self-dual gauge theory \cite{DonalSelfDual} and the non-linear sigma model \cite{CliffXYZ}.

Color-kinematics duality implies that the NNSE is invariant under the independent global symmetries,
\eq{
\textrm{color:} \quad& u_i^a \rightarrow u_i^a + f^{abc} \theta^b u_i^c \\
\textrm{kinematic:} \quad& u_i^a \rightarrow u_i^a + f_{ijk} \theta_j u_k^a \, ,
}{eq:NANS_CK_sym}
where the $\theta$ parameters are constant.  Note that the kinematic transformation is the global subgroup of diffeomorphisms, i.e.~it is {\it literally} a translation.
Without an action for the NNSE we cannot use Noether's theorem to derive the associated conserved currents.  However, it is natural to define a vector current
$J_{li} = f_{ijk} u_j^a \overset{\leftrightarrow}{\partial_l} u_k^a$
whose divergence is $\partial_l J_{li}=\frac{1}{\nu}f_{ijk}u_j^a \overset{\leftrightarrow}{\partial_0} u_k^a$
after plugging in the NNSE.
The volume integral of this quantity is the dissipation rate of kinematic charge,
\eq{
\partial_0 Q_i = \int d^3x~\partial_l J_{li} = \frac{1}{\nu} \int d^3x~f_{ijk}u_j^a \overset{\leftrightarrow}{\partial_0} u_k^a \, .
}{eq:dotQdef}
Since the integrand is a total derivative the kinematic charge is constant, $
\partial_0 Q_i=0$.  To understand this fact diagrammatically, think of $\partial_0 Q_i$ as a three-particle vertex connecting the root leg to two fluid quanta which then cascade decay into the external sources. Due to the space integral in \Eq{eq:dotQdef} the root leg is soft.  Furthermore, the $\overset{\leftrightarrow}{\partial_0}$ in \Eq{eq:dotQdef} implies that the vertex is multiplied by $\omega_1 -\omega_2 = \sum_{A_1\neq B_1} (p_{A_1}p_{B_1}) - \sum_{A_2 \neq B_2} (p_{A_2}p_{B_2})  $,
where we have used momentum conservation and the on-shell conditions.  Here $A_1,B_1$ and $A_2,B_2$ run over the decay products of the first and second fluid quantum at the vertex, respectively.   Since $\omega_1 -\omega_2$ is the difference between two inverse propagators, $\partial_0 Q_i$ simply pinches the propagators adjacent to the root leg.  Summing over all possible diagrams yields sums of triplets of kinematic numerators which vanish by the Jacobi identity.

 To implement the double copy \cite{BCJ1, BCJ2, BCJReview} we substitute all color factors with kinematic numerators.  At the level of equations of motion this is achieved by squaring the NNSE term by term to obtain the TNSE
\eq{
(\partial_0 - \nu \partial^2) u_{i\bar{i}} + \frac{1}{2}\left( u_{j\bar{j}}\partial_j \partial_{\bar{j}} u_{i\bar{i}} - \partial_j u_{i\bar{j}} \partial_{\bar{j}} u_{j\bar{i}} \right) &= J_{i\bar{i}} \, ,
}{eq:square}
which governs the dynamics of a bi-fluid velocity field $u_{i\bar{i}}$.   As with all double copies, the barred and unbarred indices of the TNSE exhibit two independent rotational invariances.  Such twofold symmetries are to be expected in any double copy \cite{CliffTwoFoldSymGR, CliffXYZ}.  Note that it is also possible to substitute the kinematic numerators for color factors to obtain the fluid analog of biadjoint scalar theory.

\medskip 

\Section{Classical Solutions.}  Lastly, we derive monopole solutions to the NNSE and the TNSE.  For the NNSE we assume an $SU(2)$ color group and a static, spherically symmetric ansatz reminiscent of the 't Hooft-Polyakov monopole \cite{tHooft, Polyakov}, $u_i^a  =f(r)   \epsilon_{aij} x_j$.
The NNSE becomes 
\eq{
f''(r)+\frac{4 f'(r)}{r}-\frac{f(r)^2}{\nu} &=0 \, ,
}{}
which admits a singular solution, $f(r) = -\frac{2\nu}{r^2}$.  For the TNSE we assume a static, spherically symmetric ansatz $u_{i \bar{i}}  =   g(r) \delta_{i \bar{i}} + h(r)  x_i x_{\bar i}$.
This yields a set of differential equations for $g$ and $h$, not detailed here, which admit a singular solution, $g(r) = 2\nu$ and $h(r) = \frac{C}{r^4}$, for an arbitrary constant $C$.  Comparing solutions side by side,
\eq{
u_i^a  =-\frac{2 \nu  \epsilon_{aij} x_j}{r^2} \qquad \textrm{and} \qquad u_{i \bar{i}}  =   2\nu \delta_{i \bar{i}} + \frac{C  x_i x_{\bar i}}{r^4}\, ,
}{eq:class_sol}
we find a structure almost identical to the Kerr-Schild double copy for monopoles and black holes \cite{KerrSchild}. Note that the classical solutions in \Eq{eq:class_sol} depend crucially on the balance between the linear and nonlinear terms in the equations of motion.  Also, the equations of motions admit non-singular solutions which can be solved for numerically but which we do not study further here.

\medskip

\Section{Conclusions.} The present work leaves numerous avenues for future inquiry.  First and foremost is the problem of turbulence and whether any insight can be gleaned from the scattering of fluid quanta at arbitrary multiplicity, e.g.~with the tools of eikonal resummation or Wilson loops \cite{Migdal:1993mb}.  Related to this is the question of whether the S-matrices for the NNSE and the TNSE exhibit an analog of the Parke-Taylor formula \cite{ParkeTaylor}.  

Second, the miraculous appearance of color-kinematics duality in the NNSE and the TNSE hints at the enticing possibility that these theories might be but a part of a larger unified web of double copy theories.  It is then natural to seek supersymmetric or stringy extensions of our results, as well as fluid analogs of the scattering equations \cite{CHY1, CHY2, CHY3, CHY4} and transmutation relations \cite{Cheung:2017ems}.

Third, given that the NNSE and TNSE exhibit color-kinematics duality off-shell, it should be possible to draw an explicit connection between the classical and amplitudes double copy.  Furthermore, it is likely that there exist other classical double copy solutions, e.g.~including spin but perhaps also relating to known solutions in fluid mechanics such as the Taylor-Green vortex.  

\medskip

\Section{Acknowledgments.}
C.C. and J.M. are supported by the DOE under grant no. DE- SC0011632 and by the Walter Burke Institute for Theoretical Physics.  We would like to thank Maria Derda, Andreas Helset, Cynthia Keeler, Julio Parra-Martinez, Ira Rothstein, and Mikhail Solon for discussions and comments on the draft.

\bibliography{scattering_NS_arxiv}

\end{document}